\documentclass{emulateapj}
\accepted{to AJ}
\usepackage{color}
\usepackage{hyperref}

\shortauthors{Matthews, Evans, \& Rupen}
\shorttitle{Radio Emission from $\delta$~Cephei}

\begin{document}
\newcommand{\ang}{\rm \AA}
\newcommand{\msun}{M$_\odot$}
\newcommand{\lsun}{L$_\odot$}
\newcommand{\days}{$d$}
\newcommand{\degree}{$^\circ$}
\newcommand{\ud}{{\rm d}}
\newcommand{\as}[2]{$#1''\,\hspace{-1.7mm}.\hspace{0mm}#2$}
\newcommand{\am}[2]{$#1'\,\hspace{-1.7mm}.\hspace{0mm}#2$}
\newcommand{\ad}[2]{$#1^{\circ}\,\hspace{-1.7mm}.\hspace{0mm}#2$}
\newcommand{\lsim}{~\rlap{$<$}{\lower 1.0ex\hbox{$\sim$}}}
\newcommand{\gsim}{~\rlap{$>$}{\lower 1.0ex\hbox{$\sim$}}}
\newcommand{\HA}{H$\alpha$}
\newcommand{\HII}{\mbox{H\,{\sc ii}}}
\newcommand{\kms}{\mbox{km s$^{-1}$}}
\newcommand{\HI}{\mbox{H\,{\sc i}}}
\newcommand{\HeI}{\mbox{He\,{\sc i}}}
\newcommand{\jks}{Jy~km~s$^{-1}$}

\title{First Detection of Radio Emission Associated with a Classical Cepheid}

\author{L. D. Matthews\altaffilmark{1}, N. R. Evans\altaffilmark{2}, 
\& M. P. Rupen\altaffilmark{3}}

\altaffiltext{1}{MIT Haystack Observatory, 99 Millstone Road, Westford, MA
  01886 USA; lmatthew@haystack.mit.edu}
\altaffiltext{2}{Center for Astrophysics $|$ Harvard \& Smithsonian, 60
  Garden Street, Cambridge, MA 02138 USA}
\altaffiltext{3}{National Research Council, Penticton, British Columbia, Canada}

\begin{abstract}
We report the detection of 15~GHz radio continuum emission associated with the
classical Cepheid variable star $\delta$~Cephei based on observations with the Karl
G. Jansky Very Large Array.  Our results 
constitute the first probable detection of radio continuum emission
from a classical Cepheid.
We observed the star at pulsation
phase $\phi\approx$0.43 (corresponding to the phase of
maximum radius and minimum temperature)
during three pulsation cycles in late
2018 and detected statistically significant emission ($>5\sigma$) during one
of the three epochs.   The observed radio
  emission appears to be variable at a $\gsim$10\% level on timescales of days to weeks.
  We also present an upper limit on the 10~GHz
flux density at pulsation phase $\phi=0.31$ from an observation in 2014.
 We discuss
 possible mechanisms that may produce the observed 15~GHz emission, but cannot make a conclusive identification
 from the present data.   The emission
  does not appear to be consistent with  originating from a close-in, late-type dwarf
companion, although this scenario cannot yet be strictly excluded.  Previous
X-ray observations have shown that $\delta$~Cephei undergoes periodic
increases in X-ray flux during pulsation phase 
$\phi\approx0.43$. The lack of radio detection in two out of three observing epochs at $\phi\approx0.43$ 
suggests that either
the radio emission is not linked with a particular pulsation phase, or else that the strength of the generated
radio emission in each pulsation cycle is variable.

\end{abstract}

\keywords{Cepheid variable stars -- Stellar atmospheres --- Radio sources}  

\section{Introduction\protect\label{intro}}
Classical Cepheid variable stars are yellow  supergiants of mass $\sim4$--$20M_{\odot}$ whose regular pulsation periods are
strongly correlated with their intrinsic luminosities (e.g., Turner 1996). This combination has led to their use as
fundamental calibrators of the
cosmic distance scale, making these stars of vital importance
to extragalactic astronomy and cosmology
(Leavitt 1908; Freedman et al. 2001; Di Benedetto 2013 and references
therein). They also play a key role in testing stellar evolution models of intermediate
mass stars (e.g., Neilson et al. 2016).
However, despite more than a century of study,
important gaps remain in our understanding of the physics
and evolution of Cepheids.

One of the most confounding puzzles is the
decades-old problem known as the ``Cepheid mass
discrepancy'': stellar mass estimates based on theoretical evolutionary
models are found to be systematically higher than masses derived observationally from
pulsation (via the mass-dependent Period-Luminosity relation) or from
orbital dynamics (e.g., Christy 1968; Pietrzy\'nski
et al. 2010). Discrepancies of
$\sim$10-15\% have persisted despite continued improvements in
models (e.g., Caputo et al. 2005; Keller \& Wood
2006; Neilson \& Langer 2012). Proposed
solutions have included extra mixing, rotation, the need for better radiative
opacities, and perhaps most importantly, mass-loss (e.g., Cox 1980; Bono
et al. 2006;
Neilson et al. 2011, 2012a, b).

If mass loss is occurring during the Cepheid evolutionary phase, this could have important
implications for the use
of classical Cepheids as distance indicators, since the presence of
circumstellar material may add scatter to inferred
luminosities in the form of extra extinction in the visible and excess
emission at infrared wavelengths (Neilson et al. 2009; 
Gallenne et al. 2013). Indeed, accounting
for these effects may be one of the keys to resolving the discrepancy
between the Hubble constant determination from Cepheids compared with
that derived from Cosmic
Microwave Background measurements (e.g., Riess et al. 2016). 
Mass loss on the instability strip would also impact other
evolutionary aspects of intermediate mass stars, 
including whether the star undergoes a blue supergiant phase
(e.g., Dohm-Palmer \& Skillman 2002; Humphreys 2010; Beasor, Davies, \& Smith 2021)
and whether the star will end its life as a white
dwarf or as a supernova. 

Despite the predictions of Cepheid mass loss, the mechanism(s)
through which Cepheids may be generating winds and driving
significant levels of mass loss (up to $\sim10^{-6}~M_{\odot}$
yr$^{-1}$) has remained an unsolved puzzle. 
Pulsationally-driven shocks have been suggested as a candidate
(e.g., Willson \& Bowen 1986; Neilson \& Lester 2008, 2009), but historically there was little
observational evidence linking pulsations with the transport
of material to distances well beyond the stellar photosphere where
mass loss can occur. However, that has begun to change.

It was recently discovered  that the classical Cepheid $\delta$~Cephei ($\delta$~Cep)
exhibits sharp ($\sim\times$4), periodic increases in
X-ray emission near pulsation phase $\phi\sim$0.43  (Engle et al. 2017). This
corresponds to
the phase just after a radially pulsating Cepheid passes through
maximum radius  and is near its temperature minimum.  The recurrent nature
of the X-ray ``bursts''  at the same phase of $\delta$~Cep's 
pulsation cycle establishes
unambiguously that they are pulsationally modulated and not linked
to an unseen companion.
Evidence of a similar phenomenon has since been seen in the classical Cepheid $\beta$~Dor (Engle et al. 2017; Evans et al. 2020).
The origin of the periodic bursts and their implication for $\delta$~Cep's
atmospheric physics are not presently understood, but the most
likely  explanation appears to be that the X-ray enhancements originate from either
flare-like
coronal activity or pulsationally-driven shocks (Engle et al. 2017; Moschou et al. 2020).
Either phenomenon could play a role in driving Cepheid mass
loss. 

 Centimeter wavelength
radio observations are sensitive to stellar emission from a variety of origins, including
free-free emission from ionized winds, chromospheres, and coronae, as well as coronal
gyrosynchrotron and gyroresonant emission
emission from active regions (e.g., G\"udel 2002 and references therein).
To gain additional insights into the origin and underlying physics of the $\delta$~Cep
X-ray bursts and to further explore their possible link with mass loss, we have used the Karl G. Jansky Very Large Array 
(VLA)\footnote{The VLA is a facility of the National Radio Astronomy
  Observatory (NRAO). NRAO is operated by Associated
  Universities, Inc. under cooperative agreement with the National
  Science Foundation.} 
to undertake radio wavelength observations during
four observing epochs, three
of which were timed to corresponded to maximum radius ($\phi\approx$0.43) during the stellar pulsation cycle.
Here we report the successful detection of emission toward the position of $\delta$~Cep during one of
these latter epochs.
We briefly discuss possible origins of
the detected radio emission and its implications for our understanding of the
previously discovered X-ray bursts.

\section{The Target: $\delta$ Cephei}
\subsection{Basic Properties\protect\label{properties}}
$\delta~$Cep is the archetype of classical Cepheid variables and is
the second closest Cepheid to the Sun (after Polaris). It is a
fundamental mode pulsator with a period of 5.366~days. Some of its additional
properties are summarized in Table~1.

$\delta$~Cep is believed to be part of a multiple system. HD~213307
(spectral type B7-8~III-IV) is a suspected 
companion at a
projected separation of $\sim40''$ to the south, and this latter star itself
may be a binary with an F0~V companion (Benedict et al. 2002). In addition, Anderson et
al. (2015) have reported evidence that $\delta$~Cep is a single-lined
spectroscopic binary. However, the putative spectroscopic companion has not yet been
detected directly. The near-infrared
interferometric study of Gallenne et al. (2016)  placed constraints
on the companion separation and spectral type of $\lsim$24~mas ($<$6.1~AU) and
later than F0V, respectively. With the availability of {\it Gaia} proper motion data, Kervella et al. (2019a, b)
further constrained the nature of the companion, narrowing the spectral type to between K3~V and M0~V.
The possible implications of a such a companion for interpreting
the radio measurements that we report here are
discussed in Section~\ref{discussion}.

\subsection{Previous Evidence of Ongoing Mass Loss\protect\label{ongoing}}
Several previous studies have provided tantalizing evidence that $\delta$~Cep is
actively losing mass at a significant rate. Based on infrared imaging observations, Marengo et al. (2010) reported the
discovery of an extended nebula and bow shock-like structure
surrounding $\delta$~Cep. Bow shocks are generally
hallmarks 
of the interaction between mass-losing stars and the ambient interstellar
medium. Therefore the presence of such a feature is strongly suggestive of a wind and ongoing mass
loss. Matthews et al. (2012) subsequently reported the discovery of a
nebula of atomic hydrogen  (detected
through its \HI\ 21~cm line emission) surrounding the stellar position. The \HI\ nebula 
spans $\sim$13$'$ (i.e., a projected size of $\sim$1~pc), and the combined properties of the nebula
and \HI\ line profile are consistent with a wind
with a mean outflow velocity of $V_{\rm out}\approx$35~\kms\ and a
mass-loss rate between $1.7\times 10^{-7}$ and $1.6\times10^{-6}~M_{\odot}$ yr$^{-1}$ after scaling to the distance adopted
in the current paper. 

Evidence of circumstellar material has also been reported on smaller
scales, closer to the photospheric surface of $\delta$~Cep. Using
near-infrared interferometry, M\'erand et al. (2006) found
emission extending to $2.4R_{\star}$ that they interpreted as evidence of circumstellar
material. Based on optical
interferometry, Nardetto et al. (2016) also found excess emission on scales of a few
stellar radii which they suggested may arise from either a circumstellar envelope or nebular
material.

\begin{deluxetable*}{llccccccccc}
\tabletypesize{\tiny}
\tablewidth{0pc}
\tablenum{1}
\tablecaption{Properties of $\delta$ Cephei}
\tablehead{
\colhead{$\alpha$} &
\colhead{$\delta$} & \colhead{Spec.}  &  \colhead{$d$} &
\colhead{$\mu_{\alpha}{\rm cos(\delta)}$} &
  \colhead{$\mu_{\delta}$}   &
\colhead{$P$} & \colhead{$M$} &
\colhead{$R_{\star}$} & \colhead{$T_{\rm eff}$}
& \colhead{log$\frac{L_{\star}}{L_{\odot}}$}  \\
\colhead{(J2000.0)} &
\colhead{(J2000.0)} & \colhead{Type}  &  \colhead{(pc)} &
\colhead{(mas yr$^{-1}$)} &
  \colhead{(mas yr$^{-1}$)}   &
\colhead{(days)} & \colhead{($M_{\odot}$)} &
\colhead{($R_{\odot}$)} & \colhead{(K)} & \colhead{} \\
  \colhead{(1)} & \colhead{(2)} & \colhead{(3)} &
\colhead{(4)} & \colhead{(5)} & \colhead{(6)} & \colhead{(7)}
& \colhead{(8)} & \colhead{(9)} & \colhead{(10)} & \colhead{(11)} }

\startdata

22 29 10.2952$\pm$0.0003 & +58 24 54.7590$\pm$0.0003 & F5Iab & 255 & 14.56$\pm$0.15 & 3.24$\pm$0.14 & 5.366 & 4.80$\pm$0.72 & 40.0 & 5960 & 3.27 
\enddata
\tablewidth{40pc}
\tabletypesize{\footnotesize}
\tablecomments{Units of right ascension are hours, minutes, and
seconds. Units of declination are degrees, arcminutes, and
arcseconds.  Coordinates and proper motions
were adopted from the third {\it Gaia} data
release (DR3)$^{*}$. Explanation of columns: (1) \& (2) right
ascension and declination (J2000.0); coordinates have been referenced to
epoch 2018.98 based on the quoted proper motion values in columns 5 and 6; (3) spectral type;
(4) adopted distance in pc (Benedict et al. 2007); (5) proper motion in right ascension,
corrected for the cosine of the declination, in mas per year; (6)
proper motion in declination in mas per year; (7) 
pulsation period in days  (Fernie et al. 1995)$^{\dagger}$; (8) mass in solar units from Kervella et al. (2019a); (9)
mean stellar radius in solar units from M\'erand et al. (2015), scaled to our adopted distance;
(10) mean effective temperature in Kelvin from Fry \& Carney (1999); (11) logarithm of stellar luminosity in solar
units, derived
from $-2.5{\rm log}(L_{\star}/L_{\odot}) = M_{V,\star}-M_{V,\odot}+{\rm BC}$,
  where the adopted solar absolute $V$ magnitude is $M_{V,\odot}$=4.73, the
  stellar absolute $V$ magnitude is taken to be $M_{V,\star}=-4.04-2.43({\rm log}P
  -1.0)$ (Evans et al. 2013), and the
  bolometric correction (BC$=-0.053$) is taken from Flower (1996). }

\tablenotetext{*}{\url{https://gea.esac.esa.int/archive/}}
\tablenotetext{$\dagger$}{The URL for the database of Fernie et al. (1995) has
      been updated since the original publication. It is currently available at
      \url{https://www.astro.utoronto.ca/DDO/research/cepheids/cepheids.html}.}

\end{deluxetable*}


%
\begin{deluxetable*}{lcllcccccccc}
\tabletypesize{\tiny}
\tablewidth{0pc}
\tablenum{2}
\tablecaption{Summary of VLA Observations}
\tablehead{
\colhead{Obs. date}  & \colhead{$\nu_{0}$} & \colhead{UT Start} & \colhead{UT Stop} & \colhead{$\phi$} &
\colhead{$N_{\rm ant}$} & 
\colhead{$t$} & \colhead{$\theta_{a}$} & \colhead{$\theta_{b}$} &
\colhead{PA} &
\colhead{$\sigma_{\rm RMS}$}     &  \colhead{$S_{\nu}$} \\
\colhead{(YYYY-MMM-DD)} & \colhead{(GHz)} & \colhead{(hh:mm:ss)} & \colhead{(hh:mm:ss)} & \colhead{} & \colhead{} & \colhead{(min)}
& \colhead{(arcsec)} & \colhead{(arcsec)} & \colhead{(deg)} &
\colhead{($\mu$Jy beam$^{-1}$)} & \colhead{($\mu$Jy)}\\
\colhead{(1)} & \colhead{(2)} & \colhead{(3)} & \colhead{(4)} & \colhead{(5)}
& \colhead{(6)} & \colhead{(7)} & \colhead{(8)} & \colhead{(9)} & \colhead{(10)} & \colhead{(11)} & \colhead{(12)}}

\startdata

2014-Oct-16 & 10.0 & 01:12:35 & 01:46:55 & 0.31 & 27 &  30.3 & 2.5 & 1.9 & +52
& 4.3 & $<$12.9 \\

2018-Nov-21 & 15.0 & 20:33:19 & 22:07:58 & 0.43 & 26 & 52.5 & 2.0 & 1.5 & +85 &
2.91 & $<$8.7 \\

2018-Nov-27 & 15.0& 05:02:04 & 06:37:38 & 0.43 & 20 & 56.3 & 2.2 & 1.3 & $-78$ & 2.93 & $<8.7$\\

2018-Dec-24 & 15.0 & 01:18:26& 02:49:50 & 0.43 & 25 & 53.8 & 1.7 & 1.4 & $-37$ &
2.76 & 15.2$\pm$2.7 \\

2018-Nov-21/27 combined & 15.0 & ... & ... & 0.43 & ...& 108.8 & 2.1 & 1.4 & $-84$ & 1.69 & $<$6.1 \\

2018 combined & 15.0 & ... & ... & 0.43 & ...& 162.6 & 1.8 & 1.4 & $-77$ & 1.69 & 7.9$^{*}\pm$1.6 

\enddata
\tabletypesize{\footnotesize}
\tablenotetext{*}{Value should be regarded as a mean over the duration of the combined data sets.}
\tablecomments{Explanation of columns: (1) observing date; (2) center observing frequency; (3) \& (4) UT
start and stop times; (5) stellar pulsation phase, computed using the emphemeris from Table~4 of Engle (2014);
(6) number of available VLA antennas; (7)
total on-source integration time; (8) dirty beam major axis;
(9) dirty beam minor axis; (10) dirty beam PA; (11) RMS
noise; (12) $\delta$~Cep flux density. Data properties are measured from images with
${\cal R}$=1 weighting (see Section~\ref{timed}). Source flux densities were derived
using Gaussian fits to the images. Quoted upper limits are 3$\sigma$. }

\end{deluxetable*}


%
\begin{deluxetable*}{lccccc}
\tabletypesize{\tiny}
\tablewidth{0pc}
\tablenum{3}
\tablecaption{Calibration Sources}
\tablehead{
\colhead{Source} & \colhead{$\alpha$(J2000.0)} &
\colhead{$\delta$(J2000.0)} & \colhead{Flux Density (Jy)} &
\colhead{$\nu$ (GHz)} & \colhead{Date (YYYY-MMM-DD)}
}

\startdata

3C48$^{a}$  & 01 37 41.2994 & +33 09 35.133 & 2.6651$^{*}$ &  10.0 & 2014-Oct-16 \\
& ...           & ...                      & 1.7842$^{*}$ & 15.0  & 2018-Nov-27\\
& ...           & ...                      & 1.7842$^{*}$ & 15.0 & 2018-Dec-24\\

3C286$^{a}$ & 13 31 08.2880 & +30 30 32.959 &  3.3580$^{*}$ & 15.0 & 2018-Nov-21\\

J2148+6107$^{b}$ & 21:48:16.0454 & +61:07:05.838 &  0.883$\pm$0.002  &  10.0  & 2014-Oct-16\\

J2250+5550$^{b}$&22 50 42.8511 & +55 50 14.581   & 0.321$\pm$0.001 & 15.0 & 2018-Nov-21\\
                & ...          & ...             & 0.260$\pm$0.001 & 15.0 & 2018-Nov-27\\
                & ...          & ...             & 0.258$\pm$0.001 & 15.0 & 2018-Dec-24\\

\enddata
\tabletypesize{\footnotesize}
\tablecomments{Units of right ascension are hours, minutes, and
seconds, and units of declination are degrees, arcminutes, and
arcseconds. $\nu$ is the frequency at which the flux density in
the fourth column was computed.}
\tablenotetext{*}{Flux densities were calculated using the
  time-dependent coefficients from Perley \& Butler (2013). For 3C48, the flux
density $S_{\nu}$ as a function of frequency was taken to be
${\rm log}(S_{\nu}) = 1.3322 - 0.7688({\rm
  log}(\nu)) - 0.1952({\rm log}(\nu))^{2} +0.0593({\rm log}(\nu))^3$, where
  $\nu_{\rm GHz}$ is the frequency expressed in GHz. For 3C286, ${\rm
  log}(S_{\nu}) = 1.2515 - 0.4605({\rm
  log}(\nu)) - 0.1715({\rm log}(\nu))^{2} +0.0336({\rm
  log}(\nu))^3$.}

\tablenotetext{a}{Primary flux calibrator and bandpass calibrator.}
\tablenotetext{b}{Gain calibrator.}

\end{deluxetable*}


\section{Observations and Data Reduction\protect\label{observations}}
\subsection{Pulsation Phase-Constrained Observations at 15 GHz\protect\label{timed}}
We obtained observations of $\delta$~Cep using the VLA during three
epochs in late 2018. Each observational epoch corresponded to a
separate pulsation cycle and was
time-constrained to commence within $\pm$15 minutes of the start of
pulsation phase $\phi$=0.43---i.e., just after the passage of the star
through maximum radius (see Table~2).  The first two observations (on November 21 and 27, respectively)
corresponded to contiguous pulsation cycles, while our third observation (on December 24)
was separated by a span of five complete pulsation cycles.

Because the nature and origin of any possible radio emission from
$\delta$~Cep were not known a priori, we chose an observing
frequency of 15~GHz (Ku band; $\lambda\approx$2.0~cm) to provide sensitivity to
emission produced by various 
possible mechanisms, both thermal and non-thermal (see Section~5.2).
Ku band also
provides relative ease in
calibrating instrumental and atmospheric phase variations.

The 2018 observations were obtained in the VLA C configuration 
(0.035--3.4~km baselines), providing a sufficiently compact array to
minimize the impact of atmospheric fluctuations on gain calibration,
but with sufficient spatial resolution
to minimize the risk of source confusion.
We used the 3-bit correlator mode that allowed three baseband pairs
(each with 2048~MHz bandwidth in dual circular polarization), for a total
observing bandwidth of $\sim$6~GHz in dual (right and left) circular polarizations. 
Each baseband contained 16 subbands, respectively divided into 128 channels of
width 1.0~MHz.

During each of the three 2-hour observing sessions, observations
of the target star were interleaved with observations of a
neighboring point source (J2250+5550)
at a projected separation of \ad{3}{9}  to provide calibration of the complex
gains. The observing sequence involved repeated cycles of $\sim$80~s on the star
and $\sim$40~s on the calibrator.  Additionally, either 3C48 or 3C286 was observed
as an absolute flux density calibrator and bandpass calibrator once
per session (see Table~3). The data dump time was 2.0~s.

Data processing was performed using the Astronomical Image Processing
System (AIPS; Greisen 2003). The archival
science data model (ASDM) format
files were loaded into AIPS using the {\sc BDFI}n
program from the Obit software package (Cotton 2008).
Antenna positions were updated to the best available values. After
flagging visibly corrupted data, a requantizer gain correction was
applied using the AIPS program {\small\sc{TYAPL}}. 
Subsequently, a fringe fit
  was performed  using a 1-minute segment
    of data on the bandpass calibrator (3C48 or 3C286) to correct the
  instrumental delays.
Bandpass calibration was performed in the standard manner, and the
absolute flux density scale was calculated by adopting
the time-dependent flux density values from Perley \& Butler (2013).

Calibration of the frequency-independent portion of the complex gains
was performed using a standard approach. First,
phase-only corrections were solved for and applied, followed by
amplitude and phase corrections. Following some additional flagging of
corrupted data, the entire calibration procedure was repeated for each
day.  Portions of the observing band were heavily
impacted by radio frequency interference (RFI) leading to overall
sensitivity losses of $\sim$10\%. As a final step, 
the data weights were optimized using the AIPS task {\small\sc{REWAY}}.

As seen in Table~3, the derived flux density of the gain calibrator
J2250+5550 is  $\sim$20\% higher on 2018 November 21 compared with
on 2018 November 27 and December 24. The 
absolute flux density scale for the
November 21 data was set using 3C286,
while the latter two days utilized 3C48, which was believed to be undergoing a flare at the time of our
observations. It is expected that this flare may contribute uncertainties of up to $\sim$10\% at 
Ku band.\footnote{\url{https://science.nrao.edu/facilities/vla/docs/manuals/oss/performance/fdscale}}
Part of the discrepancy in the J2250+5550 flux density on different days may also be the result of intrinsic
variations in the source itself.  We conservatively assume
an overall uncertainty of 20\% in our absolute flux scale.

Imaging of the fully calibrated data was performed using the AIPS {\small\sc
  IMAGR} task with a Briggs robustness parameter of ${\cal R}$=1 and a
cell size of \as{0}{28}. Initially, the data from each day were imaged
separately. A combined data set was also created from all three
days. The resulting images are shown in Figure~\ref{fig:Kuimages} and
their properties  are summarized in Table~2.

\subsection{10~GHz Observation at an Arbitrary Pulsation Phase}
In addition to the three epochs of time-constrained observations
described above, we analyzed an observation of
$\delta$~Cep obtained in 2014 under a separate program (14B-196).  The
2014 observation was obtained at a slightly lower frequency ($\nu$=10.0~GHz, X band), also using
the C configuration. This observation was executed at an
arbitrary pulsation phase ($\phi$=0.31; see Table~2) and thus does not
correspond to the pulsation phase when enhanced X-ray flux has been
seen (see Section~1). 
$\delta$~Cep was observed during two scans of $\sim$17 and $\sim$13
minutes duration, respectively, interleaved with a series of three observations of the
gain calibrator J2148+6107 (see Table~3). 

The 2014 measurement used 3-bit sampling in dual circular
polarizations. There were two independent baseband pairs, each with
a bandwidth of 2~GHz per polarization, for a total bandwidth of
4~GHz. Each baseband was divided into 16 subbands, each with 128 spectral
channels. The dump time was 3.0 seconds.

Data reduction for the 2014 observations followed the same series of steps as
the 2018 observations described in Section~\ref{timed}. Like the 2018 data, the 10~GHz observation was
significantly impacted by RFI over portions of the observing band, resulting in the need
to flag $\sim$20\% of the visibilities.

Imaging of the 2014 data was done in a similar manner to the 2018 data.
A cell size of \as{0}{3} was adopted. 
Parameters are summarized in Table~2. 

\section{Results}
\begin{figure*}
  \centering
    \hspace{0cm}
    \scalebox{0.5}{\rotatebox{0}{\includegraphics{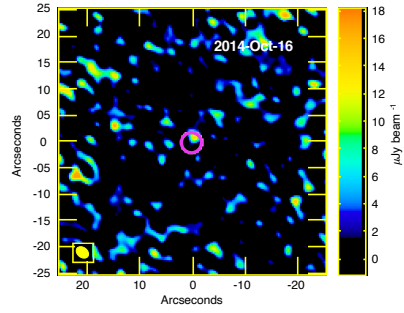}}}
     \vspace{-3cm}
\caption{Image of the $\delta$~Cep field at 10~GHz obtained with the VLA on 2014 October 16 at pulsation
  phase $\phi\approx 0.31$. The field-of-view is 50$''$.  No statistically significant emission
  was detected from $\delta$~Cep during this epoch.  The overplotted
  pink circle has a diameter of 4$''$ and is centered at the expected position of the star. The brightest positive feature visible within
  the circle has a significance of $<3\sigma$. The size of the
synthesized beam is indicated by a yellow ellipse in the lower left corner. The image properties are given in Table~2.}
\label{fig:X-band}
\end{figure*}

Figure~\ref{fig:X-band} shows a portion of the 10~GHz image from 2014 October 16, centered on the predicted 
{\it Gaia} proper motion-corrected source position of the star on this date
($\alpha_{\rm J2000}$=22$^{\rm h}$ 29$^{\rm m}$ 10.290$^{\rm s}$,
$\delta_{\rm J2000}$=+58$^{\circ}$ 24$'$ \as{54}{745}). A circle of diameter 4$''$ (approximately
twice the mean synthesized beam diameter)
  is overplotted at the source position for reference. No statistically significant emission ($>3\sigma$)
  is detected at the position of $\delta$~Cep.
  The brightest feature visible within the plotted circle has a significance
  of $\sim2.8\sigma$ and based on a Gaussian fit
  it is displaced from the predicted position of the star by  \as{-1}{55}$\pm$\as{0}{54} in
  RA and \as{1}{13}$\pm$\as{0}{34}
in DEC.
We derive a 3$\sigma$ upper limit to the $\delta$~Cep
flux density at 10~GHz of $<12.9\mu$Jy (Table~2).

\begin{figure}
\vspace{-4.5cm}
 \hspace{-2.5cm}
 \scalebox{0.75}{\rotatebox{0}{\includegraphics{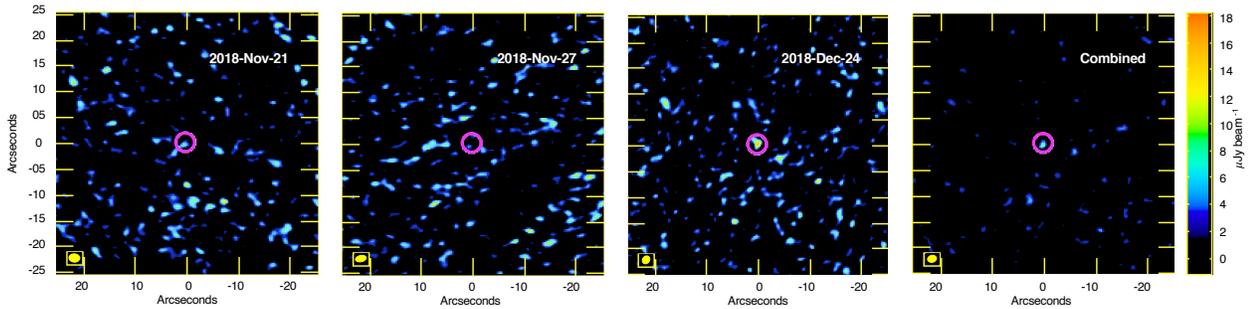}}}
 \vspace{-6cm}
\caption{Images of the  $\delta$~Cep field at 15~GHz obtained with the VLA obtained on three different dates in 2018 (left
  three images), along with a combined image based on the data from all three epochs (right-hand image). The field-of-view of each image
  is 50$''$. All observations shown were obtained
at pulsation phase $\phi\approx 0.43$.  The overplotted
pink circles are centered on the predicted proper motion-corrected position of the star and have a diameter of 4$''$. Statistically significant
emission is detected at the stellar position on December 24 (5.5$\sigma$) and in the combined image (4.7$\sigma$). The 
synthesized beam is indicated by a yellow ellipse in the lower left corner of each
  panel.  The image properties are summarized in Table~2. }
\label{fig:Kuimages}
\end{figure}

Similarly, 
during the first two observing epochs of the 2018 observations, no statistically significant
emission was detected at or near the expected position of $\delta$~Cep (Figure~\ref{fig:Kuimages}). We
place $3\sigma$ upper limits on the 15~GHz flux density of $<$8.7$\mu$Jy
on both days, and based on an image made from the combined November 21 and 27 data we derive a $3\sigma$
  upper limit of
$<6.1\mu$Jy  (Table~2). However, during the third observing epoch,
a point source with a peak flux density of $15.2\pm2.7\mu$Jy
(significance $\sim5.5\sigma$) is detected at a position consistent with the predicted (proper motion-corrected)
coordinates of $\delta$~Cep given in Table~1. Based on a Gaussian fit to the image (using AIPS task {\sc JMFIT}), the derived
position of this
feature corresponds to $\alpha_{\rm J2000} = 22^{\rm h} 29^{\rm m} 10.281^{\rm s} \pm 0.011$,
$\delta_{\rm J2000} = 58^{\circ} 24'$ \as{54}{965}$\pm$\as{0}{112}.
The position of the source thus agrees with the predicted position of $\delta$~Cep in RA to within $<1\sigma$ and in DEC to
within $<2\sigma$,
or approximately 0.1 times the FWHM of the synthesized beam.
Assuming that the detected emission arises from $\delta$~Cep and
not a spatially unresolved companion object (see Section~\ref{unseen}), this represents, to our knowledge, the
first reported probable detection of radio continuum emission associated with a Cepheid.
Dividing the December 24 data into 
time bins of various durations
we found no evidence for statistically significant variability during the observing window, implying the emission is steady
over timescales of a few hours.

Given the RMS noise levels in each of our 2018 observations (see Table~2), the source
seen in the December 24 data should have been detectable in the other two epochs if its flux density were
roughly constant over longer
timescales.
Assuming a minimum threshold for robust detection of $5\sigma$, we thus infer that the emission detected toward $\delta$~Cep
must be variable at a level of $\gsim$10\% on timescales of days or weeks. 

As shown in Figure~\ref{fig:Kuimages} and Table~2, if we average the data from the three observing epochs in 2018,
emission is again detected at the expected stellar position. A Gaussian fit yields a  source position of
$\alpha_{\rm J2000} = 22^{\rm h} 29^{\rm m} 10.299^{\rm s} \pm 0.015$, $\delta_{\rm J2000} = 58^{\circ} 24'$ \as{54}{685}$\pm$\as{0}{145},
which is consistent with the expected stellar position (Table~1) to within 1$\sigma$ uncertainties.
However, the derived flux density
(7.9$\pm1.6\mu$Jy) is smaller than that derived from the December 24 data alone, and the significance of the detection is
lower (4.7$\sigma$). These results are consistent with no 15~GHz emission being present
at the position of $\delta$~Cep during the November 21 and
November 27 observing epochs  at a level of $>3\sigma$ (see above).

Given the small angular size of the synthesized beam in our current observations (see Table~2),
the probability of a chance alignment with an extragalactic radio source is expected to be extremely small.
For example, at 1.4~GHz, Jackson (2005) estimated the number of extragalactic sources
with flux densities $\ge$10~$\mu$Jy
to be 7113 per square degree. Taking the approximation
that source counts are roughly similar across centimeter wavelength bands, this
would imply a chance probability of $\sim$0.01\% 
of a random source above that flux threshold
appearing within a given synthesized beam for our 15.0~GHz observations.
The temporal variability of the source we have
detected is also inconsistent with certain classes of extragalactic objects
(e.g., normal star-forming galaxies) which tend to be constant in time.

\section{Discussion\protect\label{discussion}}
\subsection{Previous Radio Observations of Cepheids}
There have been a handful of previous attempts to detect radio continuum
emission from classical Cepheid variables (including $\delta$~Cep) dating back several decades (Smoli\'nski et al. 1977; Welch \&
Duric 1988; Drake et al. 1991). One of the original motivations of these studies was
the detection of free-free emission from possible ionized outflows from these stars.
However, these early efforts were 10--20 times less sensitive than our recent measurements and resulted only in upper
limits. Fortunately, the continuum sensitivity of the VLA has since 
improved by more than an order of magnitude (Perley et al. 2011), enabling far
more stringent constraints on emission from stellar sources. 
Below we discuss several possible scenarios to explain the recently observed 15~GHz radio emission
associated with $\delta$~Cep. 

\subsection{Origin of the Observed Radio Emission toward $\delta$~Cep\protect\label{origin}}
\subsubsection{Photospheric Emission}
Adopting the stellar
radius, effective temperature, and distance from Table~1, one can estimate the expected thermal blackbody
emission from the stellar photosphere of $\delta$~Cep at $\nu$=15~GHz using the Rayleigh-Jeans approximation:
$S_{\rm BB}({\rm 15 GHz}) = \frac{2\nu^{2}kT_{\rm eff}\Omega}{c^{2}}\approx 1.62 \mu$Jy where $k$ is the Boltzmann constant, $\Omega$ is the solid
angle subtended by the star, and $c$ is the speed of light. This is an approximation, since stars are not perfect
blackbodies and the exact flux will depend on the radius at which a unity optical depth
($\tau\approx$1) is reached in the atmosphere for a given
wavelength (see, e.g., Drake \& Linsky 1986). However, despite these uncertainties, it is clear that
photospheric blackbody emission would be well below the
detection threshold of our current observations (see Table~2). Furthermore, in contrast with the time-variable
nature of the radio
source we have detected,
photospheric blackbody emission is not expected to vary significantly with time. 

Based on the above estimate we infer that $\lsim$20\% of the radio emission toward $\delta$~Cep originates
from  photospheric blackbody emission from the Cepheid. Since such emission is expected to be
  optically thick, this
also implies
that the radio emission that we detected on 2018 December 24
(assuming it originates from $\delta$~Cep and not a companion;
see Section~\ref{unseen}) arises predominantly from higher
regions of the atmosphere. 

\subsubsection{Free-Free Emission from a Chromosphere/Transition Region\protect\label{chromosphere}}
At $\nu$=15~GHz the emission from the quiet Sun is dominated by optically thick free-free emission from the chromosphere
and transition region (e.g., White 2004), whose plasma temperatures lie significantly above those in the photosphere.  
Previous studies of X-ray emission and optical and ultraviolet emission lines 
have established that Cepheids too can harbor hot ($10^{4}-10^{7}$~K)
plasma indicative of chromosphere-like regions
(Kraft 1957; Schmidt \& Parsons 1984; Sasselov \& Lester 1994a, b; Engle et al. 2014).
Chromospheric indicators in Cepheids are generally
found to be
time-variable  and pulsation phase-dependent, thus pulsationally-driven shocks are suspected of being a significant source of heating
(e.g., Engle et al. 2014), with possible additional heating from magnetic or acoustic wave dissipation (Sasselov \& Lester 1994b).

Adopting the stellar parameters from Table~1 and assuming $r=R_{\star}$ we estimate a mean, disk-averaged brightness
temperature for the 15~GHz emission detected on 2018 December 24 to be $\sim$55,800~K. To put this in context, such a temperature is
$\sim$5 times higher than the  mean disk-averaged brightness temperature of the quiet Sun at this frequency (Zirin et al. 1991; White 2004).
On the other hand, if we assume that the chromosphere of a typical Cepheid
extends to several tenths of a stellar radius above the photosphere (Schmidt \& Parsons 1984; Hocd\'e et al. 2020a),
the inferred mean brightness
temperature  would be reduced  (e.g.,
$T_{B}\approx 25,000$~K for a chromospheric radius of $r\sim1.5 R_{\star}$, roughly
comparable to the mean values seen in the solar chromosphere during solar maximum; White 2004).

The existence of plasma
  with temperatures of 25,000--56,000~K associated with $\delta$~Cep is consistent with the
  results of ultraviolet emission line
  measurements (Engle et al. 2014).  Thus
chromospheric free-free emission would
appear to be one plausible candidate to explain the observed radio emission from $\delta$~Cep. Further, the time-variable nature of Cepheid chromospheres may account for 
the time-varying nature of radio emission. We note, however, that in this scenario the
filling factor (and hence the true brightness temperature) of the radio-emitting plasma is highly uncertain, and that
the sources of the radio and ultraviolet emission need not be cospatial (see, e.g., White et al. 1994; Lim et al. 1998).
In the chromospheric interpretation the radio emission
would be expected to increase in brightness temperature at higher frequencies, hence future quasi-contemporaneous
observations in multiple VLA bands could provide further constraints on the viability of this
possibility. In principle the spectral index could also be estimated using  a single wide observing
  band. Unfortunately our present data have an insufficient signal-to-noise ratio to permit a meaningful constraint
  on
  the spectral index across the 6~GHz observing passband.

\subsubsection{Free-Free Emission from an Ionized Shell\protect\label{hocde}}
Hocd\'e et al. (2020b) recently postulated that Cepheids are surrounded by thin shells of ionized gas, with a characteristic
thickness of $\sim$15\%
of the stellar radius and an ionized mass of $10^{-9}$ to $10^{-7}M_{\odot}$.
The authors propose these shells as a means to explain the infrared excesses that have been observed
around a number of classical Cepheids, including $\delta$~Cep (e.g., Kervella et al. 2006; M\'erand et al. 2006;
Nardetto et al. 2016). Hocd\'e et al. (2020b) further speculate that such shells of ionized gas may be linked
to the chromospheric activity in Cepheids (see Section~\ref{chromosphere}),
although the temperatures that they propose for the shells based on radiative
transfer modeling ($T\sim$3500 to 4500~K) are
significantly lower than typical chromospheric and transition region temperatures. 

The ionized shells proposed by
Hocd\'e et al. (2020b) are expected to give rise to free-free emission at radio wavelengths. These authors
estimate flux densities for 5 Cepheids discussed in their paper to be in the range 0.01--0.1$\mu$Jy at 5~GHz. 
Although their sample does not include $\delta$~Cep, it includes V~Cen,
a Cepheid with a comparable period ($P$=5.494~days)
but a larger distance
($d$=628.7~pc). If we take V~Cen as a proxy for $\delta$~Cep and
assume that the radio emission from the shell is optically thick, the expected
flux density should scale as $S_{\nu} \propto d^{-2}\nu^{2}$. At our observing frequency of 15~GHz, we thus estimate
a predicted
flux density of $\sim0.5$--5$\mu$Jy for $\delta$~Cep based on the Hocd\'e et al. model.\footnote{Hocd\'e et al.
  (2020b)
  do not quote specific radio flux density estimates for the individual
  stars in their sample. We therefore base our scaled estimates on their quoted range.}
These values are $\sim$3--30 times smaller than
the emission that we have detected on 2018 December 24. Furthermore, 
 the cool temperatures proposed for the shells in the Hocd\'e et al. model are inconsistent with the
 disk-averaged brightness temperature of the observed radio emission if a shell
 extent of $r\sim1.15R_{\star}$ is assumed.
 This
model also assumes the shell remains static over time, in contrast with the time-variable emission that we observe.
As discussed by Hocd\'e et al. (2020b), this latter assumption is based on the lack of evidence for time-variability of the infrared excesses
around Cepheids. 

We conclude that while in principle the radio emission we observe toward $\delta$~Cep
may be consistent with a spatially extended ionized
shell, limits on the mean electron temperature from the radio emission are inconsistent with the current
``cool'' ionized shell model proposed by Hocd\'e et al. (2020b). Part of this discrepancy may result from the fact
that the models of Hocd\'e et al. do not include compression and shocks, which are expected to play a role in heating
and ionizing the gas (see also Be\u{\i}gman \& Stepanov 1981). Inclusion of these effects may also
help to reconcile the time-variable
nature of the radio emission that we observe with the predictions of the ionized shell model.

\subsubsection{Free-free Emission from a Partially Ionized Wind}
As discussed in Section~\ref{ongoing},  previous observations have provided evidence
that $\delta$~Cep is actively losing mass through a stellar wind.
Given the relatively low surface gravity of $\delta$~Cep (log~$g$=1.91 in cgs units, i.e., $\sim$0.30\% that of the Sun), its high inferred mass-loss
rate
(up to ${\dot M}\sim10^{-6}~M_{\odot}$ yr$^{-1}$),
the presence of circumstellar \HI\ 21~cm emission,  and the rather low outflow velocity estimated for the
wind ($V_{\rm out}\approx35$~\kms), 
Matthews et al. (2012) postulated that $\delta$~Cep's wind is
likely to be predominantly cool and neutral (see also Holzer \& MacGregor 1985; Glassgold \& Huggins
1983).
Nonetheless, the pulsations suspected of driving
Cepheid outflows (e.g., Willson \& Bowen 1986) are predicted to generate shocks that may propagate beyond the photosphere and
lead to partial (and time-variable) ionization of the wind (Fokin et al. 1996; Marengo et al. 2002; Nardetto et
al. 2006; Belova et al. 2014). In this case, free-free emission may be detectable at centimeter wavelengths
from the partially ionized wind.

In contrast with the case of a static chromosphere or ionized shell where flux density scales with frequency as $S_{\nu}\propto \nu^{2}$
(Sections~\ref{chromosphere} and \ref{hocde}), the frequency dependence of the free-free emission from an ionized,
expanding wind is less steep (e.g., $S_{\nu}\propto \nu^{0.6}$ in the case where the density follows an inverse square law;
Wright \& Barlow 1975; Panagia \& Felli 1975).
Using the analytic formulae provided by Drake \& Linsky (1986) we can translate our measured 15~GHz flux
density into
an estimate for the mass-loss rate through an ionized wind.
Eq.~4 from  Drake \& Linsky (1986) provides an expression to estimate
the effective size of the radio emission based on the observing frequency, measured flux density, and the wind
temperature. Assuming a canonical ionized wind temperature of $10^{4}$~K yields a radio size estimate for $\delta$~Cep
of 3.0~mas, or roughly twice
the angular diameter of its photosphere. This in turn would
imply an optically thick wind, consistent with the expectation
for cool-wind giants (Drake \& Linsky 1986).

Adopting Drake \& Linsky's
Eq.~2 for the mass-loss rate from an optically thick wind and assuming 
$V_{\rm out}\approx35$~\kms\ (see above) 
yields an estimated mass-loss rate for ionized gas of
${\dot M}^{\rm thick}_{\rm ion}\approx 4.8\times10^{-10} M_{\odot}$ yr$^{-1}$. This is several orders of
magnitude smaller than
the mass-loss
rate for $\delta$~Cep that was estimated previously from \HI\ measurements by
Matthews et al. (2012; see Section~2.2).
Taking the wind parameters from Matthews et al. (2012) and assuming this wind is the source of the observed radio
continuum emission from $\delta$~Cep, our current measurements thus imply that the
ionization fraction  must be rather low and also variable, ranging between $\sim$0.0005 (during times when radio emission
is detected) and $\lsim0.0003$ (when no radio emission is seen).

A lower limit to the expected ionization fraction in the $\delta$~Cep wind can be
estimated from stellar atmosphere models. For example, from the solar abundance models of Kurucz (1979), the ionization fraction expected
in the outer layer of the photosphere for
a low-gravity (log~$g$=2.0) star with $T_{\rm eff}=6000$~K is $\sim0.00017$, comparable to our radio-derived upper limit.
We stress, however, that there is currently considerable uncertainty in the empirically derived mass-loss rate for $\delta$~Cep, since
it depends on several poorly constrained assumptions (see discussion in Matthews et al. 2012). Based on infrared measurements, Marengo et al. (2010)
derived rates 1--2 orders of magnitude smaller than Matthews et al. (2012)---i.e., in the range $7.2\times10^{-9}$ to
$2.1\times10^{-8}M_{\odot}$ yr$^{-1}$, 
which would imply 
a higher fractional  ionization.\footnote{Values have been rescaled assuming
  a distance of 255~pc and a stellar space velocity of 10.3~\kms\ (Matthews et al. 2012).} However,  these latter
mass-loss rate estimates assumed a gas-to-dust ratio of 100 and an outflow speed of 100~\kms, respectively. As discussed by Matthews
et al. (2012), this gas-to-dust ratio
may be underestimated by an order of magnitude
for the circumstellar environments of supergiants (e.g., Skinner \& Whitmore 1988; Mauron \& Josselin 2011), while the outflow
speeds of supergiant winds are more typically a few tens of \kms, well below the stellar escape velocity (see also Holzer \& MacGregor 1985;
Judge \& Stencel 1991). These adaptations would bring the mass-loss estimates of Marengo et al. (2010) and Matthews et al. (2012) in 
much closer agreement (see Matthews et al. 2012 for discussion). We note also that these mass-loss rate
  estimates derived from
  the mass of nebular material assume a constant outflow with time. If the mass loss is instead episodic,
  this may explain why the current mass-loss rate estimated from the ionized gas appears smaller than expected.

\subsubsection{Gyroresonant Emission from Active Regions}
Engle et al. (2017) suggested that one possible origin for the period-dependent X-ray bursts observed
from $\delta$~Cep
is the occurrence of magnetic reconnection events tied to spatially compact,
heated regions covering a small fraction of the stellar surface. If present, such features
may also give rise to gyroresonant emission at radio wavelengths, analogous to the emission seen
emanating from solar active regions at frequencies
of $\sim$3 to 15~GHz (e.g., White \& Kundu 1996). More specifically, in the non-relativistic limit, low-order
harmonics ($s$=2--4)  of the gyrofrequency $\nu_{\rm B}$ are expected to give rise to
gyroresonant radio emission at frequency $\nu$, according
to: $\nu = s\nu_{\rm B} = 2.8\times10^{-3} s B$~GHz where $B$ is the magnetic field strength.

Relatively little is presently known about the magnetic fields of Cepheids. While they do possess outer convective zones,
based on their long rotational periods (e.g., $\sim$200 days for $\delta$~Cep; De Medeiros et al. 2014)
Cepheids have generally not been regarded as candidates for strong magnetic fields or
rotationally modulated magnetic activity.
However,
Barron et al. (2022a, b) recently 
reported magnetic field detections of several Cepheids based on spectropolarimetric observations---including $\delta$~Cep,
for which Barron et al. (2022a)
measured a disk-averaged longitudinal field strength $\left<B_{z}\right>=0.43\pm0.19$~G at pulsation phase
$\phi=$0.94.  

What conditions would be necessary to explain the observed radio emission from $\delta$~Cep as gyroresonant emission?
In the event that active regions
are present, then the local magnetic strength within such regions would be expected to be significantly higher than
the disk-averaged value. 
Assuming, for example, emission arising from the third harmonic ($s$=3),
the production of radio emission at $\nu$=15~GHz would require a local
magnetic field strength of $\sim$1800~G, i.e., more than three orders of magnitude higher than the disk-averaged field.
The strength of the resulting radio emission would then depend on the covering fraction of active regions.
Following O'Gorman et al. (2017; see also Drake et al. 1993; Villadsen et al. 2014) we can estimate the filling factor of
such regions, $f_{\rm 15GHz}$, by assuming that the observed 15~GHz brightness temperature $T_{\rm 15GHz}$ can be expressed as

\begin{equation}
T_{\rm 15GHz} = (1 - f_{\rm 15GHz})T_{\rm disk} + f_{\rm 15GHz}T_{\rm act}
\end{equation}

\noindent where  $T_{\rm disk}$ is the mean stellar disk brightness temperature and $T_{\rm act}$ is the brightness
temperature of a typical active region. We assume that the stellar disk is uniform except at the discrete  locations
where a coronal active region is present. Taking $T_{\rm 15GHz}=$55,800~K (Section~\ref{chromosphere}), $T_{\rm disk}=$5690~K (Table~1), and
$T_{\rm act}\approx 2\times 10^{7}$~K (based on the X-ray measurements of Engle et al. (2017) during maximum radius) we find
$f_{\rm 15GHz}\approx$0.0025. Assuming instead $T_{\rm act}\approx 10^{6}$~K 
  (comparable to the brightness temperature of
gyroresont emission above active regions on the Sun; White 2004) yields $f_{\rm 15GHz}\approx$0.05.
 The latter is comparable to the coverage factor of 2\% estimated by Moschou et al. (2020) as necessary
  to explain $\delta$~Cep's X-ray emission
via discrete magnetic reconnection events.

Additional observations and modeling will be needed to establish whether
  the magnetic properties of $\delta$~Cep support this interpretation. 
  Barron et al. (2022a) noted that the magnetic field amplitude and morphology of $\delta$~Cep are somewhat peculiar and reminiscent of
  those of certain Am stars, including Sirius A. However, in the case of Sirius A, the magnetic field does not appear to play a
  role in generating its observed radio emission. For Sirius~A the radio flux between wavelengths of $\sim$0.4 to 9.0~mm is found
  to be constant in
  time and to have a mean, disk-averaged
  brightness temperature consistently below that of the photosphere (White et al. 2018, 2019). These characteristics
  are well reproduced by an LTE model of Sirius's photosphere extended to radio wavelengths and are consistent with free-free rather than gyromagnetic
  emission
  (White et al. 2018).

\subsubsection{Emission from a Close-in Companion\protect\label{unseen}}
As discussed in Section~\ref{properties}, there is evidence that $\delta$~Cep 
has at least two stellar companions. One of these, HD~213307 (a B7-8~III-IV star with its own possible F0~V companion), lies at a projected
separation of $\sim40''$ and is therefore spatially well-resolved from $\delta$~Cep in our 2018 VLA
observations. No radio statistically significant radio emission was detected at the position of
HD~213307 during any of our VLA observing epochs. Using the
combined 15~GHz image
from our 2018 observations (see Table~3), we place a 3$\sigma$ upper
limit on the quiescent radio emission from HD~213307 to be $<5.1~\mu$Jy.

The second candidate companion to $\delta$~Cep (believed to be a K3 to M0 dwarf; see Section~\ref{properties})
lies projected angular separation of $\lsim$24~mas (Anderson et al. 2015).
This is only a small fraction of the synthesized
beam in our VLA observations (see Table~3), and thus we cannot
spatially resolve $\delta$~Cep from this putative companion. Could it
be the source of the observed time-varying radio emission?

The coronae of cool, late-type dwarfs frequently give rise to both {\em quiescent} radio
emission and {\em flares} at centimeter wavelengths (e.g., G\"udel 2006). Given the
relatively young age of $\delta$~Cep ($\sim$80~Myr; Matthews et
al. 2012), its close companion would likely be a rapidly rotating,
magnetically active star, similar to those found in the Pleiades
(G\"udel 2002; Engle et al.  2017), making it a plausible radio
emitter. However, the characteristics of the observed radio emission do not seem to fit
neatly with such an interpretation.

The  radio emission detected toward
$\delta$~Cep on 2018 December 24 does not exhibit  any
of the typical hallmarks of flaring or bursting emission from a cool dwarf
(see, e.g., Bastian et al. 1990; G\"udel 2002; Osten \& Bastian 2006, 2008);
for example, we do not see any temporally
discrete features (either narrow- or broad-band) attributable to bursts or flares visible in the dynamic
spectra (cf. Osten \& Bastian 2008;
Lynch, Mutel, \& G\"udel 2015; Route \& Wolszczan 2016) and we find no evidence of
temporal modulation of the radio emission over the course of the December 24
observation. The detected emission also does not exhibit any
detectable level of circular polarization, as is commonly seen in both
flaring and quiescent emission from cool dwarfs at $\nu\lsim$10~GHz (e.g., Osten \& Bastian 2008; Villadsen \& Hallinan 2019),
although we note that
detectable circular polarization appears
to be less common
at $\nu\approx$15~GHz compared with radio frequencies $\lsim$10~GHz (e.g., White et al. 1994; Osten et al. 2006; Fichtinger et al. 2017).

In contrast to bursts or flares,
quiescent emission from cool, low-mass dwarfs is expected to be broadband and vary slowly on timescales or hours or
days, consistent with the source that we observe.
Typical radio luminosities of such stars are $10^{12}$--$10^{16}$ erg s$^{-1}$ Hz$^{-1}$ (G\"udel 2006), also consistent with our data:
the observed flux density of the source we detect on 2018
December 24 corresponds to a radio luminosity $L_{\rm R}\approx 1.2\times10^{15}$ erg s$^{-1}$ Hz$^{-1}$.
However, other properties of the data do not seem to fit this picture.

Assuming that the radio emission from a cool dwarf is optically thick at 15~GHz (e.g., White et al. 1994),
the inferred radio brightness temperature $T_{B}$ equals the mean electron temperature ($T_{e}$), while in the optically
thin case $T_{B}$ is a lower limit to $T_{e}$ (e.g., G\"udel et al. 1995).
Adopting
fiducial radii $r\sim 0.8R_{\odot}$ for a K3~V star and  $\sim0.6R_{\odot}$ for an
M0~V star, respectively (Lang 1991), the 15.2~$\mu$Jy
flux density we measure at 15~GHz translates to a disk-averaged brightness
temperature of $T_{B}\gsim(1.4-2.5)\times10^{8}$~K---significantly hotter than a typical corona. If instead
the emission arises from material extending
to $r\sim$2--3$R_{\star}$ with a near-unity filling factor, temperature estimates
($T_{B}\sim(2-6)\times10^{7}$~K) are more in
line with typical coronal temperatures of active K and M dwarfs as derived from X-ray observations (White et al. 1994; Giampapa et al. 1996).  However,
persistent plasma temperatures of $>10^{7}$~K are inconsistent
with X-ray measurements of the $\delta$~Cep system.

The X-ray observations presented by
Engle et al. (2017) did not have  the spatial resolution to separate $\delta$~Cep from a source
with an angular separation of $<$24~mas, and therefore these measurements effectively
provide an upper limit to the X-ray luminosity and temperature of 
any such companion. However, the plasma temperatures inferred from the quiescent X-ray emission
  toward $\delta$~Cep by Engle et al. ($T_{e}\sim10^{6}$~K) are more than
an order of magnitude smaller than the brightness temperature that we derive
from the radio emission under the assumption that
it arises from a region with a radius comparable to that of a cool dwarf corona.
The only times plasma temperatures are observed to exceed $10^{7}$~K in the environs of $\delta$~Cep are during the periodic bursts
at pulsation phase $\phi\approx$0.43. However, Engle et al. (2017) have shown that these temperatures
appear to  be linked with the atmosphere of $\delta$~Cep itself, not a dwarf companion.

The radio/X-ray temperature discrepancy could potentially be reconciled if the X-ray-emitting plasma and
the electrons responsible for the radio emission at 15~GHz are not co-spatial. However, 
the dwarf companion to $\delta$~Cep would then be atypical of cool dwarfs in exhibiting a disk-averaged brightness temperature at
radio wavelengths that is an order of magnitude higher than than plasma temperatures inferred from X-ray measurements.
Indeed, for a sample of K and M dwarfs observed at 15~GHz, White et al. (1994) found the stars to be systematically {\it underluminous}
in the radio compared with a model in which the 15~GHz radio emission arises from optically thick gyroresonant emission with a plasma
temperature equal to that inferred from X-ray measurements (see also Osten et al. 2006).

Based on these considerations, we conclude that while a cool dwarf companion cannot yet be ruled out as the origin
of the radio emission associated with $\delta$~Cep, it does not appear to be the most likely candidate.
Additional epochs radio observations of the star, both near maximum radius
($\phi\approx$0.43)
and at other (arbitrary) pulsation phases are needed to obtain additional constraints on this possibility.
For example, radio emission that is present only near specific pulsation phases would effectively rule out
the companion as the origin.

\section{Summary} 
We have presented multi-epoch centimeter wavelength continuum observations of the classical Cepheid $\delta$~Cep
obtained with the VLA. The star was undetected in a 10~GHz observation
obtained at pulsation phase $\phi$=0.31 in 2014 October. During  late 2018 we
obtained additional observations of the star
at 15~GHz and a pulsation phase $\phi\approx$0.43 during three pulsation cycles (two of which were contiguous). The latter 
phase corresponds to maximum radius (and minimum temperature)
for the star during its 5.366 day pulsation cycle and represents the phase where periodic X-ray bursts
have been previously reported. During one of the three epochs (2018 December 24) we detected statistically
significant radio emission ($>5\sigma$) consistent with the position of $\delta$~Cep. This represents the first
probable detection of radio emission from a Cepheid variable.  The non-detection during two other
15~GHz observing epochs implies the emission is variable at a level of $\gsim$10\% on timescales of days or weeks. 
  This also suggests
  that the production of radio emission is not tied to a specific pulsation phase, unless the strength of the radio
emission produced varies significantly from cycle to cycle.
Possible origins for the detected emission include free-free emission from a chromosphere or ionized circumstellar shell,
free-free emission from an expanding, partially ionized wind, or gyroresonant emission from localized active regions. 
The properties of the radio emission do not appear to be consistent
with arising from a close-in, late-type (K or M) dwarf companion, although  this possibility cannot yet be fully excluded.
Follow-up radio monitoring of $\delta$~Cep, both during and
outside of the phase of maximum radius, will be necessary to further constrain the origin of the radio emission and to
determine whether it is correlated with pulsation
phase.

\acknowledgements
We thank the referee for a careful reading of the manuscript and J. Drake for helpful discussion.
LDM was supported by award AST-2107681 from the National Science Foundation.
Support to NRE
was provided from the {\it Chandra} X-Ray Center NASA contract NAS8-03060.
The observations presented here were part of NRAO programs AM1304
(VLA/14B-196) and AM1550 (VLA/18B-005).
This research has made use of the
VizieR catalogue access tool, 
CDS, Strasbourg, France (DOI: 10.26093/cds/vizier), the SIMBAD database,
operated at CDS, Strasbourg, France,  and the
data from the European Space Agency (ESA) mission {\it Gaia} ({\url{https://www.cosmos.esa.int/gaia}}),
processed by the {\it Gaia} Data Processing and Analysis Consortium (DPAC, {\url{https://www.cosmos.esa.int/web/gaia/dpac/consortium}}).
Funding for the DPAC has been provided by national institutions, in particular the institutions participating in the {\it Gaia}
Multilateral Agreement.



%

%

\end{document}